\documentclass[12pt]{article}
\begin{document}

\title{High Energy Effects of Noncommutative Spacetime Geometry}
\author{B.G. Sidharth\footnote{birlasc@gmail.com}, B.M. Birla Science Centre,\\ Adarsh Nagar, Hyderabad - 500 063, India\\
\\Abhishek Das\footnote{parbihtih3@gmail.com}, B.M. Birla Science Centre,\\ Adarsh Nagar, Hyderabad - 500 063, India\\}
\date{}
\maketitle
\begin{abstract}
In this paper, we endeavour to obtain a modified form of the
Foldy-Wouthuysen and Cini-Toushek transformations by resorting to
the noncommutative nature of space-time geometry, starting from the
Klein-Gordon equation. Also, we obtain a shift in the energy levels
due to noncommutativity and from these results a limit for the
Lorentz factor in the ultra-relativistic case has been derived in conformity with observations.
\end{abstract}
\maketitle

\section{Introduction}

The noncommutative feature of space-time geometry is a topic of great interest. There is a vast amount of literature
existing on this subject \cite{Michael,Meljanac,Chaichian,Louis}. Particularly, the author Sidharth has used noncommutativity
to provide a feasible interpretation for several phenomena \cite{c,db,non}. The objective of the current paper is to further
explore this unique and intrinsic nature of space-time, which was first introduced by Snyder \cite{Snyder}. \\
Particularly, we apply this noncommutativity to the Klein-Gordon equation and endeavour to modify it. The second section deals
with the modification of the Klein-Gordon equation by innovating a parameter representing the noncommutative feature. In the
third section we modify the Foldy-Wouthuysen and the Cini-Toushek transformations that represent the low energy and high
energy scenario respectively. In the fourth section, we investigate further regarding the aforesaid parameter of noncommutativity
and in that course we find some novel results concerning the Lorentz factor.

\section{Modified Klein-Gordon equation}

As the author Sidharth has mentioned in several papers
\cite{bgs1,bgs2} and several papers therein, the consideration of
complex time ($it$) leads to the Minkowski space-time formalism
while the ordinary time coordinate ($t$) leads to the compact four
space representing the {\it zitterbewegung} region. This effect was
noticed by Dirac himself when he came up with his equation: the
rapidly oscillationg solutions, apparently unphysical. Dirac's
explanation was that our physical measurements are never
instantaneous but rather spread over a small interval-- it turns out
to be the Compton time \cite{diracpqm}. Zitterbewegung has been
studied a lot over the years, notably by Huang, Hestenes, Kaiser and
other scholars \cite{huang,hestenes,kaiser}. More recently Sidharth
has re-examined it in the light of the Feshbach Villars formulation
\cite{feshbach,uhef,uhep}. To put it simply the four component Dirac
wave function can be written as
$$\psi = \left(\begin{array}{ll}
\chi \\ \phi\end{array}\right)$$ where $\chi$ and $\phi$ are each
two component spinors. $\chi$ are the so called "high energy"
spinors and $\phi$ the low energy ones. The former become pronounced at high energies and latter at lower energies. So
the Dirac four spinor divides spacetime into two broad regions-- the high energy region where $\chi$ dominates and the usual
low energy region where $\phi$ dominates.\\
We begin with the following complexified identification of time.\\
\begin{equation}
t \longmapsto \alpha t + \beta it^{\prime} \label{a}
\end{equation}
where, $\alpha$ and $\beta$ are parameters that represent the scale below and above the Compton length respectively.
When one considers phenomena below the Compton scale, $\beta = 0$, and have the ordinary time coordinate, $t$. Again,
when one considers phenomena above the Compton scale, $\alpha = 0$, and we have the complex time coordinate, $it^{\prime}$.\\
Thus, relation (\ref{a}) represents a region which is the juncture between the compact four space and the non-compact
Minkwoski space. The metrics for the two different regions are respectively\\
\[x^{2} + y^{2} + z^{2} + c^{2}t^{2}\]\\
and\\
\[x^{2} + y^{2} + z^{2} - c^{2}t^{\prime2}\]
Now, for the juncture region we write\\
\begin{equation}
{\rm d}s^{2} = x^{2} + y^{2} + z^{2} + tt^{*}
\end{equation}
This yields\\
\begin{equation}
{\rm d}s^{2} = x^{2} + y^{2} + z^{2} + \alpha c^{2}t^{2} - \beta c^{2}t^{\prime2}
\end{equation}
This represents the critical region that is the boundary of the two regions. Ostensibly, this is the region of the Compton
length. Now, the d'Alembertian operator as\\
\[\frac{\partial^{2}}{\partial x_{\mu}\partial x_{\nu}} = \nabla^{2} - \frac{1}{c^{2}}\frac{\partial^{2}}{\partial t^{2}}\]\\

Again, previously \cite{bgs3} we have shown that due to the noncommutative nature of space-time one can arrive at the
following relation\\

\begin{equation}
\frac{\partial^{2}}{\partial x_{\mu}\partial x_{\nu}} = [1 - \frac{\epsilon \beta(l^2)}{\delta x_{\nu}\delta x_{\mu}}]\frac{\partial^{2}}{\partial x_{\mu}\partial x_{\nu}} \label{b}
\end{equation}
where, $\beta(l^{2})$ is a matrix that bears the signature of a noncommutative space-time and $l$ is the Compton length.
Anyway, we shall use relation (\ref{b}) which is essentially a transformation on account of noncommutativity, in case of
the Klein-gordon equation which can be written as\\
\begin{equation}
\frac{\partial^{2}\psi}{\partial x_{\mu}\partial x_{\nu}} + \mu^{2}\psi = 0 \label{c}
\end{equation}
where $\psi = \psi(x,t)$ is the wavefunction and $\mu$ ($=\frac{mc}{\hbar}$) is the mass-related term. Again,
considering the transformation relation (\ref{b}) for the d'Alembertian and interchanging the indices ($\mu \leftrightarrow \nu$)
we can also write for (\ref{c})\\
\begin{equation}
[1 - \frac{\epsilon \beta(l^2)}{\delta x_{\nu}\delta x_{\mu}}]\frac{\partial^{2}\psi}{\partial x_{\mu}\partial x_{\nu}} =  \mu^{2}\psi \label{p}
\end{equation}
Now, writing\\
\[[1 - \frac{\epsilon \beta(l^2)}{\delta x_{\nu}\delta x_{\mu}}] = \zeta\]\\
and\\
\[\mu^{\prime} = \frac{\mu}{\zeta}\]\\
we have finally the modified Klein-Gordon equation as\\
\begin{equation}
\frac{\partial^{2}\psi}{\partial x_{\mu}\partial x_{\nu}} + \mu^{\prime 2}\psi = 0 \label{d}
\end{equation}

Here, $\zeta$ is nearly equal to 1, since the $\beta(l^{2})$ is
ostensibly infinitesimal. The equation (\ref{d}) relates the
noncommutative feature of space-time through the matrix
$\beta(l^{2})$ with the mass parameter ($\mu$). But
Cf.ref.\cite{bgsijtp,uof} for a slightly different approach. One can
infer that the generation of mass is due to the noncommutative
space-time. This is because we see from
equation (\ref{p}), $[1 - \frac{\epsilon \beta(l^2)}{\delta x_{\nu}\delta x_{\mu}}]\frac{\partial^{2}}{\partial x_{\mu}\partial x_{\nu}}$, is an operator that operates on the wavefunction and produces mass-related term $\mu$. This shows that the noncommutative feature of space-time can be indeed interesting when taken into consideration. We shall see of this more in the subsequent section.\\
Ostensibly, if the noncommutative nature of space-time is neglected then the parameter $\beta(\l^{2})$ is 0,
which leads to $\zeta = 1$ and we have the usual Klein-Gordon formulation.\\

\section{The modified transformations for high energy and low energy scenario}

In the previous section we have derived a modified form of the Klein-Gordon equation, the modification itself arising
from noncommutativity. Here, we shall derive modified forms of the Foldy-Wouthuysen \cite{Bjorken,Foldy,Pong,Yuri} and
the Cini-Toushek transformations \cite{Pong,Cini}. Now, the modified Klein-Gordon equation (\ref{d}) can also be written as\\
\begin{equation}
(\partial_{\mu}\partial_{\nu} + \mu^{\prime2})\psi = 0 \label{e}
\end{equation}
This can also be written as\\
\begin{equation}
(i\gamma_{\mu}\partial_{\mu} + \mu^{\prime})(-i\gamma_{\nu}\partial_{\nu} + \mu^{\prime})\psi = 0 \label{q}
\end{equation}
From (\ref{q}), as we know, one can infer\\
\[(\gamma_{\mu}p_{\mu} + \mu^{\prime})\psi = 0\]\\
or\\
\[(-\gamma_{\mu}p_{\mu} + \mu^{\prime})\psi = 0\]\\
where, we have taken $\hbar = c = 1$ and $p_{\mu} = i\partial_{\mu}$. Again, in presence of an electromagnetic
interaction these two equations can be rewritten as\\
\begin{equation}
[\gamma_{\mu}(p_{\mu} - eA_{\mu}) + \mu^{\prime}]\psi = 0 \label{x}
\end{equation}
and
\begin{equation}
[\gamma_{\mu}(-p_{\mu} + eA_{\mu}) + \mu^{\prime}]\psi = 0 \label{y}
\end{equation}
It is obvious that if equation (\ref{x}) represents a particle of mass 'm' and charge 'e' then equation (\ref{y})
represents the antiparticle with mass 'm' and charge '-e'. Therefore, as we see the two equations that derive from the
Klein-Gordon equation (\ref{e}) correspond to a matter-antimatter asymmetry. Now, without any interaction the
aforementioned equations can also be written respectively as
\begin{equation}
({\bf {\alpha.p}} + \beta\frac{m}{\zeta_{1}})\psi = i\frac{\partial\psi}{\partial t} \label{f}
\end{equation}
and
\begin{equation}
({\bf {\alpha.p}} - \beta\frac{m}{\zeta_{2}})\psi = i\frac{\partial\psi}{\partial t} \label{g}
\end{equation}
respectively, where, ${\bf \alpha}$ and $\beta$ are the usual matrices. Here, apparently we have distinguished
between the $\zeta$'s for the two equations (\ref{f}) and (\ref{g}). The rationale for this is the fact that these two
distinct equations represent a particle and an antiparticle. Interestingly, we shall find out later that the former
corresponds to the low energy case and the latter to the high energy case. Now, let us consider a unitary transformation
for the equation (\ref{f}) \\
\[U = e^{is}\],\\
\[\psi^{\prime} = e^{is}\psi\]\\
such that we have\\
\begin{eqnarray*}
i\frac{\partial\psi}{\partial t} &=& e^{is}H\psi \\
&=& e^{is}He^{-is}\psi^{\prime} \\
&=& H^{\prime}\psi^{\prime}
\end{eqnarray*}
where, $H$ is the usual Dirac Hamiltonian. As we know \cite{Bjorken}, such a choice of transformation is given by\\
\begin{equation}
e^{is} = e^{\beta{\bf {\alpha.p}}\theta({\bf p})} = \cos p\theta + \beta\frac{{\bf {\alpha.p}}}{p}\sin p\theta \label{h}
\end{equation}
Thus, the transformed Hamiltonian is given as\\
\begin{eqnarray*}
H^{\prime} &=&  (\cos p\theta + \beta\frac{{\bf {\alpha.p}}}{p}\sin p\theta)(({\bf {\alpha.p}} + \beta\frac{m}{\zeta_{1}}))(\cos p\theta - \beta\frac{{\bf {\alpha.p}}}{p}\sin p\theta) \\
&=& {\bf {\alpha.p}}(\cos 2p\theta - \beta\frac{m}{\zeta_{1} p}\sin 2p\theta) + \beta(\frac{m}{\zeta_{1}}\cos 2p\theta + p\sin 2p\theta)
\end{eqnarray*}
Putting \\
\begin{equation}
\tan 2p\theta = \frac{\zeta_{1} p}{m} \label{i}
\end{equation}
in the first term of the last line we obtain\\
\[H^{\prime} = \beta(\frac{m}{\zeta_{1}}\cos 2p\theta + p\sin 2p\theta)\]
Now, squaring both sides, using (\ref{i}) and after some rearranging we obtain the transformed Hamiltonian as\\
\begin{equation}
H^{\prime} = \frac{\beta}{\zeta_{1}}\sqrt{m^{2} + \zeta_{1}^{2}p^{2}} \label{j}
\end{equation}
where, we see that the effect of noncommutativity is included in the new transformed Hamiltonian, in the form of
$\zeta_{1}$. Of course, if $\beta(l^{2}) = 0$, then we have $\zeta_{1} = 1$ and we obtain the known transformed
Hamiltonian \cite{Bjorken} as\\
\[H^{\prime} = \beta\sqrt{m^{2} + p^{2}}\]\\
Incidentally, we have also found the modified unitary transformation as\\
\begin{equation}
U_{L} = e^{\beta{\bf {\alpha.p}}\theta({\bf p})} = \exp[\frac{1}{2}\beta\alpha\tan^{-1}(\frac{\zeta_{1} p}{m})]
\end{equation}
where, the subscript $L$ refers to the low energy scenario. Manifestly, this is the modification of the Foldy-Wouthuysen
transformation \cite{Foldy,Pong}. Thus, we see that taking into consideration the noncommutative nature of space-time one
gets new effects. Next, we shall consider the equation (\ref{g}) and find out if it leads to the high energy scenario.
We consider a similar type of unitary transformation as before, namely relation (\ref{h}). Therefore,we have the transformed
Hamiltonian as\\
\begin{eqnarray*}
H^{\prime} &=&  (\cos p\theta + \beta\frac{{\bf {\alpha.p}}}{p}\sin p\theta)(({\bf {\alpha.p}} - \beta\frac{m}{\zeta_{2}}))(\cos p\theta - \beta\frac{{\bf {\alpha.p}}}{p}\sin p\theta) \\
&=& {\bf {\alpha.p}}(\cos 2p\theta + \beta\frac{m}{\zeta_{2} p}\sin 2p\theta) + \beta(\frac{m}{\zeta_{2}}\cos 2p\theta - p\sin 2p\theta)
\end{eqnarray*}
Now, putting
\begin{equation}
\tan 2p\theta = \frac{m}{\zeta_{2} p} \label{k}
\end{equation}
in the second term of the last line we obtain \\
\[H^{\prime} = {\bf {\alpha.p}}(\cos 2p\theta + \beta\frac{m}{\zeta_{2} p}\sin 2p\theta)\]\\
Squaring both sides, using (\ref{k}) and rearranging the terms we would obtain\\
\begin{equation}
H^{\prime} = \frac{\alpha}{\zeta_{2}} \sqrt{m^{2} + \zeta_{2}^{2}p^{2}} \label{l}
\end{equation}
which is the new transformed Hamiltonian for the high energy scenario, considering the noncommutative effects. As usual,
for $\beta(l^{2}) = 0$, we have $\zeta_{2} = 1$ and the usual Hamiltonian\\
\[H^{\prime} = \alpha\sqrt{m^{2} + p^{2}}\]\\
Thus, we have obtained the following unitary transformation
\begin{equation}
U_{H} = e^{\beta{\bf {\alpha.p}}\theta({\bf p})} = \exp[\frac{1}{2}\beta\alpha\tan^{-1}(\frac{m}{\zeta_{2} p})]
\end{equation}
where, the subscript $H$ refers to the high energy scenario and relation (\ref{l}) is the modified Cini-Toushek transformation.
Ostensibly, we have corroborated the fact that the Klein-Gordon formulation is the combination of both the Foldy-Wouthuysen
and the Cini-Toushek formulation, where the former refers to the low energy case and the latter to the high energy case.
Besides, let us take a look at the transformed Hamiltonians that we have derived. For the low and high energy scenarios
we have respectively relations (\ref{j}) and (\ref{l}). Thus, from the modified Foldy-Wouthuysen transformation we can derive\\
\begin{equation}
H_{L}^{\prime} = U_{L}H_{L}U_{L}^{-1} = \beta E^{\prime}_{L} \label{m}
\end{equation}
where, $E^{\prime}_{L}$ denotes the modified energy levels for the low energy scenario ($m^{2}\gg p^{2}$) given as\\
\[E^{\prime}_{L} = \frac{1}{\zeta_{1}}\sqrt{m^{2} + \zeta_{1}^{2}p^{2}}\]\\
Again, from the modified Cini-Toushek transformation we have\\
\begin{equation}
H_{H}^{\prime} = U_{H}H_{H}U_{H}^{-1} = \alpha E^{\prime}_{H} \label{n}
\end{equation}
where, $E^{\prime}_{H}$ denotes the modified energy levels for the high energy scenario ($m^{2}\ll p^{2}$) given as\\
\[E^{\prime}_{H} = \frac{1}{\zeta_{2}}\sqrt{m^{2} + \zeta_{2}^{2}p^{2}}\]\\
The transformed Hamiltonians (\ref{m}) and (\ref{n}) are the same as the conventional ones \cite{Pong}, except for the
fact that the noncommutative feature of space-time has been included in them which has culminated the modification
of the energy levels. \\

\section{The parameter $\zeta$}
Now, the modified Hamiltonians (\ref{j}) and (\ref{l}) give the modified energy levels. Apparently, there is a shift from the
known values of the energy levels which should be observed in experiments. This fact can be correlated to the modified
mass-energy relation that has been studied in our previous works \cite{bgs4,bgs5}. For a general case, we can write\\
\begin{equation}
\frac{1}{\zeta}\sqrt{m^{2} + \zeta^{2}p^{2}} = \sqrt{m^{2} + p^{2} - \lambda^{2}l^{2}p^{4}} \label{v}
\end{equation}
where, $\lambda \approx -10^{-3}$ is a constant whose value we had previously found \cite{bgs4} and $l$ is the generally the Compton length ($l = \frac{\hbar}{mc}$). But here, due to relativistic considerations we consider it as the De Broglie length ($l = \frac{\hbar}{p}$) of the particle.
Incidentally, we had shown in the said paper that the constant $\lambda$ is related to the electron gyromagnetic ratio and the
Schwinger correction terms by the relation\\
\begin{equation}
g = 2[1 + \frac{\alpha}{2\pi} + f(\alpha)] = 2[1 - \lambda]\
\end{equation}
where, $\alpha$ is the fine structure constant and $f(\alpha)$ consists of higher orders of $\alpha$. From the above equation
one can easily infer that \\
\[\lambda \approx -\frac{\alpha}{2\pi}\]\\
neglecting the higher order terms consisting $\alpha$. This result also helped in explaining the {\it GZK cutoff} and
the {\it Lamb shift} phenomenon in two previous papers \cite{bgs5,bgs6}. Essentially, $|\lambda| \approx \frac{\alpha}{2\pi}$,
is the reduced fine structure constant. However, squaring both sides of (\ref{v}) and after some rearranging we derive the
following result for the parameter $\zeta$\\
\begin{equation}
\zeta = (1 - \epsilon^{2})^{-\frac{1}{2}} \label{z}
\end{equation}
where, $\epsilon = \frac{\lambda l p^{2}}{m}$, in terms of natural units ($\hbar = c = 1$). Particularly, we know that
the De Broglie length ($l$) is given as \\
\[{l} = \frac{\hbar}{p}\]\\
Considering ordinary units we have \\
\[\epsilon = \frac{\lambda lc}{\hbar} \frac{p^{2}}{mc^{2}}\]\\
Using the previous relation for $l$ and $p = mc$ the value of $\epsilon$ is\\
\begin{eqnarray*}
\epsilon &=& {\lambda c} \frac{mc}{mc^{2}} \\
&=& \lambda
\end{eqnarray*}
Now, if we hadn't considered $p = mc$ then we would have obtained\\
\begin{equation}
\epsilon = \lambda\frac{p}{mc}
\end{equation}
Thus, we would have the general value of $\zeta$ as\\
\begin{equation}
\zeta = \frac{1}{\sqrt{1 - \lambda^{2}\frac{p^{2}}{m^{2}c^{2}}}} \label{u}
\end{equation}
Now, from equation (\ref{u}) we can infer three possible cases (with $\lambda = -10^{-3}$), as follows\\
\\
1)~For, $p = mc$ we would have\\
\[\zeta \approx 1.0000005\]\\

2)~For, $p \ll mc$ (non-relativistic scenario) we have $\zeta$ nearly equal to 1 but\\
\[\zeta > 1 \]\\

3)~For, the case $p \gg mc$ (ultra-relativistic scenario) we must have\\
\begin{equation}
(\frac{p}{mc})_{\max} < 10^{3} \label{r}
\end{equation}
which is a critical value, in the absence of any interactions. If, $(\frac{p}{mc})_{\max} \geq 10^{3}$ then we would
have either an infinite or an imaginary value of $\zeta$ which would make equations (\ref{j}) and (\ref{l}) unrealistic and unphysical. Now, from special relativity we have the following relation\\
\[P = \gamma mv\]\\
where, $\gamma$ is the Lorentz factor and $v$ is the velocity of the particle under consideration. Thus, we may write\\
\[(\frac{p}{mc})_{\max} = (\frac{\gamma v}{c})_{\max} < 1000\]\\
giving\\
\[\frac{1}{\sqrt{1 - (\frac{v}{c})_{max}^{2}}}(\frac{v}{c})_{max} < 1000\]\\
From here, we would obtain \\
\begin{equation}
(\frac{v}{c})_{\max} < 0.9999995
\end{equation}
and the corresponding Lorentz factor as\\
\begin{equation}
(\gamma)_{\max} < 10^{6} \label{s}
\end{equation}
Thus, we can infer that for a spin-1/2 particle obeying (\ref{f}) or
(\ref{g}), at the ultra relativistic limit the Lorentz factor and
the factor $\frac{v}{c}$, both have an upper bound, in the absence
of any field or interaction. This upper bound is borne out of several observations
\cite{Sivaram,Swadesh,Kumar,Giannios,Xiao}. Our approach takes into account the
noncommutative nature of space-time geometry. In theoretical terms,
exceeding the critical value of (\ref{s}) will yield un-physical
phenomena. Also, as we can see, the inequality (\ref{r}) poses a
strict upper bound to the value of $\frac{p}{mc}$ or
$\frac{pc}{mc^{2}}$. This is the rationale for the parameter $\zeta$
which we had introduced in the first section.\\
When the noncommutative nature of space-time is taken into
consideration we have such feasible results. Ostensibly, neglecting
this noncommutativity takes us back to the known scenarios and
results. But, the limit to the Lorentz factor is a fruitful
derivation since it might provide further insights into
ultra-relativistic phenomena for particles obeying the relations
(\ref{f}) and (\ref{g}), where the former concurs with the low
energy scenario (Foldy-Wouthuysen case) and the latter with the high
energy scenario (Cini-Toushek case).\\
In simple terms, the non-commutative nature of spacetime which
manifests itself at ultra high energies, prohibits a runaway Lorentz
Factor $\gamma$.

\section{Discussions}

In the light of the approach considered in this paper we see that the noncommutative feature of space-time plays an important role in the understanding of several phenomena. Particularly, this inherent noncommutativity puts a restriction on the Lorentz factor. This intuition can be extended to achieve further interesting results. Let us consider the Lorentz factor\\
\[\gamma = \frac{1}{\sqrt{1 - (\frac{v}{c})^{2}}}\]\\
Differentiating both sides with respect to time we can get the acceleration as\\
\begin{equation}
a = \frac{{\rm d}v}{{\rm d}t} = \frac{c}{\sqrt{1 - \frac{1}{\gamma^{2}}}}\times \frac{1}{\gamma^{3}}\frac{{\rm d}\gamma}{{\rm d}t} \label{t}
\end{equation}
If we take the limiting value of the Lorentz factor as proposed by us, to be\\
\[\gamma \approx 10^{6}\]\\
and write\\
\[\frac{{\rm d}\gamma}{{\rm d}t} = \Delta\]
then from (\ref{t}) we get the acceleration as \\
\[a \approx 1.0000005 \times 10^{-8} \Delta ~cm/s^{2}\]\\
where, $\Delta$ is small since $\gamma$ is bounded. Thus, we can simply drop the $\Delta$ factor and finally wirte\\
\begin{equation}
a < 1.0000005 \times 10^{-8}~ cm/s^{2} \label{o}
\end{equation}
This is interesting as it is almost exactly of the order of the acceleration produced by the Cosmological Constant \cite{Smolin} which is given by $\frac{c^{2}}{R}$ (where, $R$ is the radius of the universe). In an altogether different approach, the bound in (\ref{o}) was derived earlier by Sidharth \cite{bgs7,bgs8,bgs9}. Also, this result corresponds to the anomalous acceleration of the Pioneer 10 and 11 \cite{Smolin,Anderson} and the MOND theory \cite{mil1}. Here, we have basically substantiated the fact that the noncommutative nature of space-time is in splendid accord with the results mentioned in the aforesaid papers. \\

\end{document}